\documentclass[letterpaper, 10 pt, conference]{ieeeconf}  

\IEEEoverridecommandlockouts                              

\usepackage{graphics} 
\usepackage{epsfig} 
\usepackage{times} 
\usepackage{amsmath} 
\usepackage{xcolor}

\usepackage{mathrsfs}
\usepackage{amssymb}
\usepackage{cite}
\usepackage{dsfont}
\usepackage{xcolor}
\usepackage{mathtools}
\usepackage{mathrsfs}
\usepackage{subfig}
\usepackage{balance}
\usepackage{float}
\usepackage{flushend}
\usepackage{bm}         
\usepackage{mathtools}

\usepackage{amsthm}
\usepackage{tcolorbox}
\usepackage[framemethod=TikZ]{mdframed}
\usepackage{mdframed}
\usepackage{thmtools} 
\usepackage{algorithm}
\usepackage{booktabs}
\usepackage{algpseudocode}
\theoremstyle{plain}
\newtheorem{theorem}{Theorem}

\newtheorem{problem}{Problem} 

\theoremstyle{definition}

\theoremstyle{remark}

\newmdtheoremenv[linecolor=white, backgroundcolor=lightgray!15, innertopmargin=5pt, innerbottommargin=5pt, skipabove=10pt, skipbelow=10pt]{objective}{\textbf{Objective}}
\newmdtheoremenv[linecolor=white, backgroundcolor=lightgray!15, innertopmargin=5pt, innerbottommargin=5pt, skipabove=10pt, skipbelow=10pt]{lemma}{\textbf{Lemma}}

\newcommand{\ctrlseq}{\mathbf{u}}

\newcommand{\state}[0]{x}
\newcommand{\norm}[1]{\left\lVert#1\right\rVert}

\title{\LARGE \bf
Safe and Performant Controller Synthesis using Gradient-based Model Predictive Control and Control Barrier Functions
}

\author{Aditya Singh*, Aastha Mishra*, Manan Tayal, Shishir Kolathaya, and Pushpak Jagtap
\thanks{
}
\thanks{This work was supported by the AI \& Robotics Technology Park (ARTPARK) at IISc.}
\thanks{All the authors belong to Cyber-Physical Systems, Indian Institute of Science (IISc), Bengaluru.
{\{adityasingh, aasthamishra,  manantayal, shishirk, pushpak\}@iisc.ac.in}
}%
\thanks{* denotes equal contribution.
}%
}

\begin{document}

\maketitle
\thispagestyle{empty}
\pagestyle{empty}


\begin{abstract}
Ensuring both performance and safety is critical for autonomous systems operating in real-world environments. While safety filters such as Control Barrier Functions (CBFs) enforce constraints by modifying nominal controllers in real time, they can become overly conservative when the nominal policy lacks safety awareness. Conversely, solving State-Constrained Optimal Control Problems (SC-OCPs) via dynamic programming offers formal guarantees but is intractable in high-dimensional systems. In this work, we propose a novel two-stage framework that combines gradient-based Model Predictive Control (MPC) with CBF-based safety filtering for co-optimizing safety and performance. In the first stage, we relax safety constraints as penalties in the cost function, enabling fast optimization via gradient-based methods. This step improves scalability and avoids feasibility issues associated with hard constraints. In the second stage, we modify the resulting controller using a CBF-based Quadratic Program (CBF-QP), which enforces hard safety constraints with minimal deviation from the reference. Our approach yields controllers that are both performant and provably safe. We validate the proposed framework on two case studies, showcasing its ability to synthesize scalable, safe, and high-performance controllers for complex, high-dimensional autonomous systems.
\end{abstract}

\section{Introduction}
\label{introduction}
Autonomous systems are increasingly deployed in diverse fields, including autonomous driving, robotic manipulation, aerospace, and industrial automation. A key challenge in designing controllers for such systems lies in the dual mandate of ensuring both \textit{performance} and \textit{safety}. Performance pertains to the system's ability to accomplish tasks efficiently and within resource constraints such as time, energy, or actuation limits. Safety, on the other hand, involves adherence to state and input constraints to avoid hazardous behavior or system failures. These objectives are often at odds, necessitating careful co-design. For example, in drone-based delivery systems, minimizing delivery time improves efficiency, yet aggressive maneuvers to save time may risk collisions or stability violations, especially in cluttered or dynamic environments. Therefore, synthesizing control policies that jointly optimize performance while providing safety guarantees is a fundamental requirement for the reliable deployment of autonomous systems.

A class of methods involves \textit{safety filtering} \cite{safety_filter, 10266799}, which ensures constraint satisfaction by modifying control outputs in real-time. Methods such as Control Barrier Function (CBF)-based quadratic programs (QP) \cite{Ames_2017} and Hamilton-Jacobi (HJ) Reachability filters \cite{8263977, 10665911} act as corrective layers on top of a (potentially unsafe) nominal controller, making minimal interventions to enforce safety constraints. However, if the reference controller does not account for safety in any manner, these filters can lead to myopic or overly conservative behaviors. An alternative and more rigorous strategy commonly adopted in the controls community is to formulate the problem as a State-Constrained Optimal Control Problem (SC-OCP) and solve it via dynamic programming~\cite{altarovici2013general, Hao2024csl}. Despite its theoretical elegance, this method suffers from the curse of dimensionality, making it computationally intractable for systems with more than five dimensions.

To address these limitations, we propose a two-stage framework that bridges safety filtering and SC-OCP approaches. In the first stage, we reformulate the SC-OCP as an online nonlinear constrained Model Predictive Control (MPC) problem~\cite{wang2025}. The safety constraints are relaxed and incorporated into the cost function as penalty terms, enabling the design of control policies that prioritize task performance while encouraging safety~\cite{srinivasan2020}. This relaxation eliminates hard nonlinear constraints, mitigating feasibility issues commonly encountered in nonlinear constrained optimization. The resulting unconstrained MPC problem is then solved using gradient-based optimization, which provides greater scalability and better solutions compared to sample-based techniques~\cite{pmlr-v120-bharadhwaj20a, 6862353, 9992498} such as Model Predictive Path Integral (MPPI)~\cite{8558663, 10161511}, particularly in high-dimensional systems.

In the second stage, the solution obtained from the relaxed MPC problem is used as the reference input to a Control Barrier Function-based Quadratic Program (CBF-QP). Since the reference controller already incorporates a notion of safety, the resulting filtered controller avoids the conservatism typically induced by CBF-QP post-processing. This two-step approach yields policies that are both provably safe and minimally deviating from the task-optimal trajectory. By leveraging scalable gradient-based optimization and a structure that separates performance and safety objectives, our framework effectively synthesizes high-performance, safe controllers suitable for complex, high-dimensional systems.

To summarize, the key contributions of this work are:
\begin{itemize}
    \item We introduce a novel two-stage control synthesis framework that combines gradient-based Model Predictive Control (MPC) with Control Barrier Functions (CBFs) to design controllers that are both safe and high-performing.
    
    \item By utilizing gradient-based optimization rather than sampling-based approaches, our method achieves improved computational efficiency and scalability, making it well-suited for high-dimensional systems.
    
    \item The integration of a CBF-based Quadratic Program (CBF-QP) provides formal safety guarantees, ensuring that the resulting control policies are provably safe.
    
    \item Through two case studies on unicycle and quadrotor navigation, respectively, we demonstrate the capability of the proposed framework to co-optimize safety and performance, while maintaining scalability to more complex dynamical systems.
\end{itemize}
 
\nocite{tayal2024semi, singh2025exactbc,tayal2025a, jagtap2020formal}

\section{Problem Formulation}
\label{prob_for}
Consider a control-affine system with state $x \in \mathcal{X} \subseteq \mathbb{R}^n$ and control input $u \in \mathcal{U} \subseteq \mathbb{R}^m$, governed by the dynamics $\dot{x} = f(x) + g(x)u$, where $f: \mathbb{R}^n \to \mathbb{R}^n$ and $g: \mathbb{R}^n \to \mathbb{R}^{n \times m}$ are differentiable functions.

Let $\mathcal{F} \subseteq \mathcal{X}$ denote the failure set, representing unsafe states. The system's performance is evaluated via the cost functional:
\begin{equation}
    J(t,x(t), \ctrlseq) = \int_{s=t}^{T} r(x(s)) \, ds + \phi(x(T)),
\end{equation}
where $r: \mathcal{X} \to \mathbb{R}_{\geq 0}$ and $\phi: \mathcal{X} \to \mathbb{R}_{\geq 0}$ are non-negative, Lipschitz continuous functions representing the running and terminal costs, respectively. The control signal $\ctrlseq:[t,T) \rightarrow \mathcal{U}$ drives the system. The objective is to synthesize an optimal policy $\pi^*: [t, T) \times \mathcal{X} \to \mathcal{U}$ that minimizes $J$ while ensuring the system state remains outside the failure set $\mathcal{F}$ over the time horizon $[t, T]$.

To achieve the stated objective, the first step is to encode the safety constraint via a function $l: \mathbb{R}^n \to \mathbb{R}$ such that $\mathcal{F}:= \{x \in \mathcal{X} \mid l(x) \leq 0\}$. Using these notations, the control synthesis problem can then be cast as the following State-Constrained Optimal Control Problem (SC-OCP):

\begin{problem}[State-Constrained Optimal Control Problem]\label{prob:1}
\begin{equation}\label{eq: SC-OCP}
    \begin{aligned}
    \inf_{\ctrlseq} J(t, x(t), \ctrlseq) &= \int_t^{T} r(x(s))\, ds + \phi(x(T)) \\
    \text{s.t.} \quad & \dot{x} = f(x) + g(x)u, \\
                      & l(x(s)) > 0 \quad \forall s \in [t, T]
    \end{aligned}
\end{equation}
\end{problem}

This SC-OCP enhances the system's performance by minimizing the cost, while maintaining system safety through the state constraint, $l(x) > 0$, ensuring that the system avoids the failure set, $\mathcal{F}$. Thus, the policy, $\pi^*$, derived from the solution of this SC-OCP co-optimizes safety and performance. 

\subsection{MPC Reformulation}
To solve the SC-OCP in \ref{prob:1}, we recast the problem within a Model Predictive Control (MPC) framework. Specifically, Problem~\ref{prob:1} is solved over a shorter prediction horizon $h$, the first control input of the resulting sequence is applied, and the problem is re-solved at the next time step using the updated system state. This closed-loop approach improves robustness against model uncertainties and unforeseen disturbances during execution.

To implement this MPC-based strategy for solving Problem~\ref{prob:1}, we first discretize the system dynamics and the SC-OCP. Let $\mathbf{x}$ and $\mathbf{u}$ denote the discrete-time state and control trajectories, respectively. The resulting discrete-time SC-OCP is given by:

\begin{problem}[Discrete-Time SC-OCP]\label{prob:2}
\begin{equation}
\begin{aligned}
 \min_{\mathbf{u}} \quad J(x, \mathbf{u}) &= \sum_{k=1}^{K} r(\mathbf{x}(k), \mathbf{u}(k)) + \phi(\mathbf{x}(K))\\
\text{s.t.}~ \mathbf{x}(k+1) &= f_d(\mathbf{x}(k)) + g_d(\mathbf{x}(k))\mathbf{u}(k), \quad \forall k \in \mathbb{N}, \\
l(\mathbf{x}(k)) &> 0, \quad \forall k \in \{1, 2, \ldots, K\}
\end{aligned}
\end{equation}
\end{problem}

Here, $f_d$ and $g_d$ denote the discrete-time counterparts of the system dynamics, assumed to be differentiable, analogous to $f$ and $g$ in Problem~\ref{prob:1}. Assuming a planning horizon of $h$, the MPC formulation to solve Problem~\ref{prob:2} at time step $j$ can be formalized as follows:

\begin{problem}[MPC Reformulation of the SC-OCP]\label{prob:3}
\begin{equation}
\begin{aligned}
\min_{\mathbf{u}} \quad &\sum_{k=j}^{j+h} r(\mathbf{x}(k), \mathbf{u}(k)) + \phi(\mathbf{x}(j + h)) \\
\text{s.t.} \quad \mathbf{x}(k+1) &= f_d(\mathbf{x}(k)) + g_d(\mathbf{x}(k))\mathbf{u}(k), 
~\forall k \in \mathbb{N}, \\
l(\mathbf{x}(k)) &> 0, ~ \forall k \in \{j, j+1, \ldots, j+h\}
\end{aligned}
\end{equation}
\end{problem}

Building on this premise, we formulate the central objective of this work:
\begin{objective}
\label{obj: Main_obj}
   Our goal is to develop a scalable and computationally efficient framework for solving Problem~\ref{prob:3}, which can be further leveraged to synthesize an optimal policy $\pi^*: [t, T) \times \mathcal{X} \to \mathcal{U}$ that minimizes the cost $J$ while guaranteeing that the system state avoids the failure set $\mathcal{F}$ throughout the time horizon $[t, T]$.
\end{objective}

\section{Methodology}
\label{methodology}
As discussed in Section~\ref{prob_for}, the SC-OCP formulation for jointly optimizing safety and performance entails solving the nonlinear MPC problem in Problem~\ref{prob:3} at each time step $j$ over a prediction horizon $h$. This optimization is challenging due to the complexity of system dynamics and safety constraints. Two main classes of planning approaches are commonly employed: sampling-based methods (e.g., MPPI) and gradient-based techniques. While sampling methods evaluate random action sequences, they often perform poorly in high-dimensional settings due to their reliance on scalar cost signals and lack of gradient information. In contrast, gradient-based methods leverage differentiable dynamics to obtain full cost gradients, providing more informative feedback, improved convergence rates, and hence, enhanced performance as action dimensionality increases. Furthermore, unlike sampling approaches, which face exponential growth in computational cost with dimensionality, gradient-based methods scale almost linearly. Accordingly, we adopt a gradient-based planner to solve the MPC problem in Problem~\ref{prob:3}, ensuring better performance and scalability. The following subsections describe our approach in detail.

\subsection{Gradient-Based Planning}
\begin{table*}[t]
\centering
\begin{tabular}{lccc@{\hspace{1cm}}lccc}
\toprule
\multicolumn{4}{c}{\textbf{Obstacle Avoidance}} & \multicolumn{4}{c}{\textbf{Quadrotor}} \\
\cmidrule(r){1-4} \cmidrule(l){5-8}
\textbf{Method} & \textbf{Cost} & \textbf{Safety Rate (\%)} & \textbf{Computation Time (s)} & & \textbf{Cost} & \textbf{Safety Rate (\%)} & \textbf{Computation Time (s)} \\
\midrule
GMPC           & 1.417  & 85   & 0.14 & & 1.036  & 88   & 0.3 \\
MPPI          & 2.085  & 85   & 0.24 &  & 2.366  & 87   & 0.7 \\
MPPI-CBF    & 2.071  & 100  & 0.20 &  & 2.302  & 100  & 0.7 \\
\textbf{GMPC-CBF (Ours)}     & 1.403  & 100  &  0.14 & &  1.093  & 100  & 0.3 \\
\bottomrule
\end{tabular}
\caption{Comparative analysis of all evaluated methods based on the defined evaluation metrics}
\label{comp table}
\end{table*}
The explicit inclusion of the safety constraint renders Problem~\ref{prob:3} challenging to solve directly. To address this, we aim to remove the hard safety constraint while retaining the core objective of the problem. This is accomplished by reformulating the safety requirement as a soft constraint, incorporated into the cost function through an appropriate penalty term. This leads to the following revised optimization problem formulation:

\begin{problem}[MPC Formulation with Safety as Soft Constraint]\label{prob:4}
\begin{equation}
\begin{aligned}
\min_{\mathbf{u}} \quad C(x, \mathbf{u}) = \sum_{k=j}^{j+h}~\Big( &r(\mathbf{x}(k), \mathbf{u}(k)) +
\phi(\mathbf{x}(j + h)) \\ + &\lambda \max\{0, -l(\mathbf{x}(k)) + \delta\} \Big) \\
\text{s.t.} \quad \mathbf{x}(k+1) = f_d(\mathbf{x}(k)) &+ g_d(\mathbf{x}(k))\mathbf{u}(k), 
~\forall k \in \mathbb{N}, \\
\end{aligned}
\end{equation}
\end{problem}

Intuitively, the trade-off parameter $\lambda~(> 0)$ modulates the balance between safety and performance: larger values of $\lambda$ strongly penalize even minor safety violations, emphasizing safety, while smaller values lead to more performance-oriented (and potentially aggressive) behavior, possibly at the expense of safety. The scalar $\delta > 0$ is introduced to enforce a strict inequality, effectively inflating the safety constraint to enhance robustness against potential safety violations. Furthermore, as seen in Problem~\ref{prob:4}, the original state constraints have been relaxed and incorporated as soft penalties in the cost function, yielding a more tractable optimization problem.

To solve the above optimization problem efficiently, we employ a Quasi-Newton method using the Limited-memory Broyden–Fletcher–Goldfarb–Shanno (L-BFGS) algorithm~\cite{L-BFGS}. The update rule is given by:
\begin{equation}
    \mathbf{u}(k)_{i+1} = \mathbf{u}(k)_i - \alpha_i H_i \nabla C(\mathbf{x}(k),\mathbf{u}(k)_i),
\end{equation}
where $\alpha_i$ is a step size obtained through line search, $H_i$ is an approximation of the inverse Hessian matrix, and $C$ is the cost function. L-BFGS is particularly well-suited for large-scale optimization problems due to its ability to approximate curvature information without explicitly storing or computing the full Hessian, thereby achieving low memory overhead and improved convergence properties compared to gradient descent. By leveraging second-order information implicitly, L-BFGS enhances optimization efficiency while maintaining scalability, making it ideal for real-time or high-dimensional control tasks.

While this formulation accounts for the system's safety requirements by incorporating them as soft constraints, such a treatment merely promotes safety rather than guaranteeing it. As a result, the resulting policy may still exhibit unsafe behavior, posing a significant limitation for deployment in safety-critical systems. To address this limitation, we employ a safety filtering mechanism based on Control Barrier Functions (CBFs), which enforces hard safety constraints on the controller. The specifics of the CBF-based safety filter are presented in the subsequent subsection. 

\subsection{Safe and Performant Controller Synthesis using CBFs}

\begin{algorithm}[t]
\caption{Gradient-based Planning with CBF-QP Safety Filter}
\label{alg: GMPC_CBF}
\begin{algorithmic}[1]
\State \textbf{Input:} Initial state $x_0$, goal position $x_{goal}$, horizon $H$, number of steps $T$
\State \textbf{Parameters:} Dynamics $f$ and $g$, cost $C$
\For{$t = 0$ to $T-1$}
        \State \textbf{Solve MPC:}
        \State $\mathbf{x_0} \gets x_t$
        \State Initialize $u_{0:H-1}$
        \State Optimize $u_{0:H-1}$ using L-BFGS on cost:
        \State \hskip1em $J = \sum_{k=0}^{H-1} C(\mathbf{x_k}, u_k)$, 
        \State \hskip1em $\mathbf{x_{k+1}} = \mathbf{x_{k}} + (f(\mathbf{x_k}) + g(\mathbf{x_k})u_k)\Delta{t}$
        \State $u_{\text{nom}} \gets u_0$
        \State \textbf{Apply CBF-QP Safety Filter:}
        \State $u^* \gets \text{Solve CBF-QP from \eqref{eq: MPC_QP}}$ 
        \State $x_{t+1} \gets x_t + (f(x_t) + g(x_t)u^*)\Delta{t}$
\EndFor
\end{algorithmic}
\end{algorithm}

The gradient-based planner synthesizes a controller that encourages system performance while satisfying nominal safety constraints. However, it does not offer formal safety guarantees. To enforce safety without significantly degrading performance, this controller can be minimally modified using the Control Barrier Function-based Quadratic Program (CBF-QP)~\cite{ames2014control}. Before presenting the CBF-QP formulation, we briefly review Control Barrier Functions (CBFs).

CBFs~\cite{ames2014control, Ames_2017, ames2019control} are used to construct controllers that ensure forward invariance of a designated safe set. The initial step in constructing a Control Barrier Function (CBF) involves defining a continuously differentiable function $h: \mathcal{X} \to \mathbb{R}$, where the \textit{super-level set} of $h$ corresponds to the safe region $\mathcal{C}$. This leads to the following representation:
\begin{equation}
    \begin{aligned}
    \label{eq:setc1}
    \mathcal{C} = \{\state \in \mathcal{X} : h(\state) \geq 0\}, \quad
    \mathcal{X} \setminus \mathcal{C} = \{\state \in \mathcal{X} : h(\state) < 0\} \\
     \text{Int}(\mathcal{C}) = \{\state \in \mathcal{X} : h(\state) > 0\}, \quad
    \partial\mathcal{C} = \{\state \in \mathcal{X} : h(\state) = 0\}
    \end{aligned}
\end{equation}
where, $\text{Int}(\mathcal{C})$ and $\partial\mathcal{C}$ represent the interior and boundary of $\mathcal{C}$ respectively. Furthermore, the function $h$ qualifies as a valid CBF if it satisfies the following theorem:
\begin{theorem}[Control Barrier Function (CBF)]
\label{theorem: CBF definition}
Consider a control-affine system $\dot x = f(x) + g(x)u$. Given a set $\mathcal{C}$ defined by \eqref{eq:setc1} with $\frac{\partial h}{\partial \state}(\state) \neq 0$ for all $\state \in \partial \mathcal{C}$, the function $h$ is a Control Barrier Function if there exists an extended class-$\mathcal{K}$ function $\kappa$ such that for all $\state \in \mathcal{X}$,
\begin{equation}
\label{eq: lie_derivative}
    \sup_{u \in \mathcal{U}} \left[\mathcal{L}_f h(\state) + \mathcal{L}_g h(\state)u + \kappa(h(\state))\right] \geq 0,
\end{equation}
where $\mathcal{L}_f h = \frac{\partial h}{\partial \state}f$ and $\mathcal{L}_g h = \frac{\partial h}{\partial \state}g$.
\end{theorem}

As shown in~\cite{Ames_2017, ames2019control}, any Lipschitz continuous controller $\mu(x)$ satisfying \eqref{eq: lie_derivative} ensures safety if $x(0) \in \mathcal{C}$ and guarantees convergence to $\mathcal{C}$ if $x(0) \notin \mathcal{C}$.

Based on this premise, the CBF-QP~\cite{ames2014control} minimally modifies a reference controller $u_{\mathrm{ref}}$ using a valid CBF $h$ to produce a safe control input, and is formulated as:
\begin{equation}
\begin{aligned}
\label{eq: CBF_QP}
u^*(x) &= \underset{u \in \mathcal{U}}{\arg\min}~ \norm{u - u_{\mathrm{ref}}}^2 \\
\text{s.t.} \quad & \mathcal{L}_f h(x) + \mathcal{L}_g h(x)u + \kappa(h(x)) \geq 0.
\end{aligned}
\end{equation}

By choosing the gradient-based controller $u_{\mathrm{mpc}}$ as the reference input, we obtain the final formulation:
\begin{equation}
\begin{aligned}
\label{eq: MPC_QP}
\mathbf{u}^*(x) &= \underset{u \in \mathcal{U}}{\arg\min}~ \norm{u - u_{\mathrm{mpc}}}^2 \\
\text{s.t.} \quad & \mathcal{L}_f h(x) + \mathcal{L}_g h(x)u + \kappa(h(x)) \geq 0.
\end{aligned}
\end{equation}
where $\mathbf{u}^*$ provides a safe and performant control action for the task and safety specifications of Problem~\ref{prob:3}. The steps of our proposed approach are outlined in Algorithm~\ref{alg: GMPC_CBF}.


\section{Results and Discussion}
\label{Results}

The goal of this paper is to demonstrate the efficacy of the proposed method in co-optimizing performance and safety. To this end, we evaluate our approach against baseline methods using the following three metrics: (1)~\textbf{Cumulative Cost}: This metric represents the total cost  $\sum_{k=1}^{K} r(\mathbf{x}(k), \mathbf{u}(k)) + \phi(\mathbf{x}(K))$, accumulated by a policy over the safe trajectories. (2)~\textbf{Safety Rate}: This metric represents the percentage of trajectories that remain within the safe set, i.e., never enter the failure set $\mathcal{F}$ during execution. (3)~\textbf{Computation Time}: This metric reflects the average time required to solve the MPC problem at each time step.

\textbf{Baselines and Compute Device Details:} We evaluate our proposed approach against a sample-based baseline, denoted as MPPI-CBF~\cite{rabiee2024guaranteedsafemppicompositecontrol}, wherein the reference controller for the CBF is designed using MPPI instead of gradient-based planning. This comparison enables us to assess the impact of employing gradient-based MPC for solving the underlying optimization problem. As part of the ablation study, we also compare against MPPI and gradient-based MPC (GMPC) methods that do not incorporate CBFs, in order to investigate the effect of the CBF-based safety filtering. In both these baseline cases, the planner directly solves Problem~\ref{prob:4} without any subsequent CBF-based filtering.
Finally, all experiments were performed on a machine with an 11th Gen Intel Core i9-11900K @ 3.50GHz (16 cores) CPU and 128GB RAM.

\subsection{Unicycle Robot Navigation with Collision Avoidance}

\begin{figure}[h]
    \centering
    \includegraphics[width=1\linewidth]{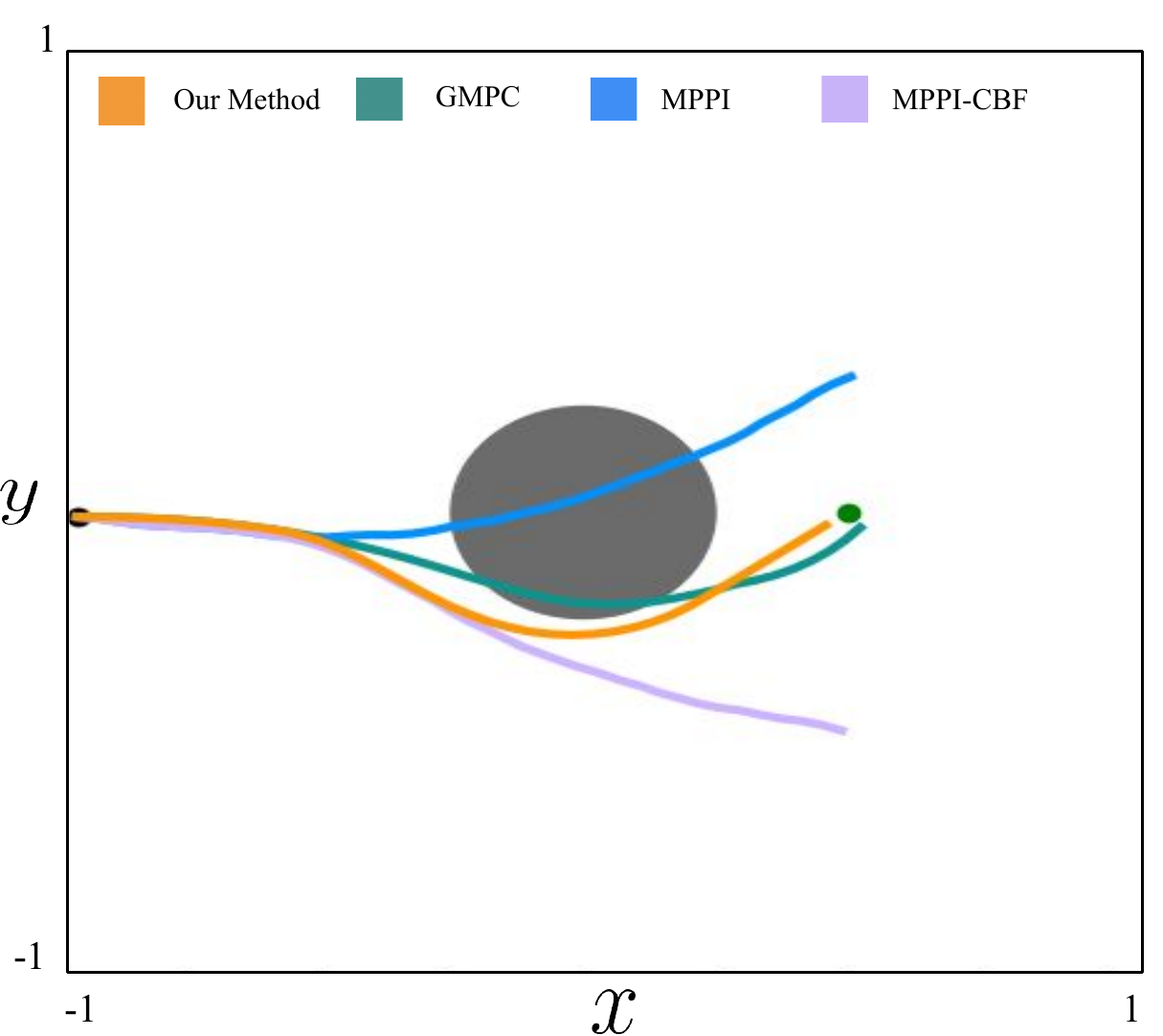}
    \caption{Trajectories from a common initial state is shown, with dark grey circles representing obstacles and the green dot indicating the goal at $[0.5, 0]^T$. Our method is the \textbf{only one} that \textbf{successfully approaches the goal} while \textbf{adhering to safety constraints}.}
    \label{fig:oa plots}
    \vspace{-1.5em}
\end{figure}

\begin{figure*}[t]
    \centering    \includegraphics[width=1.0\linewidth]{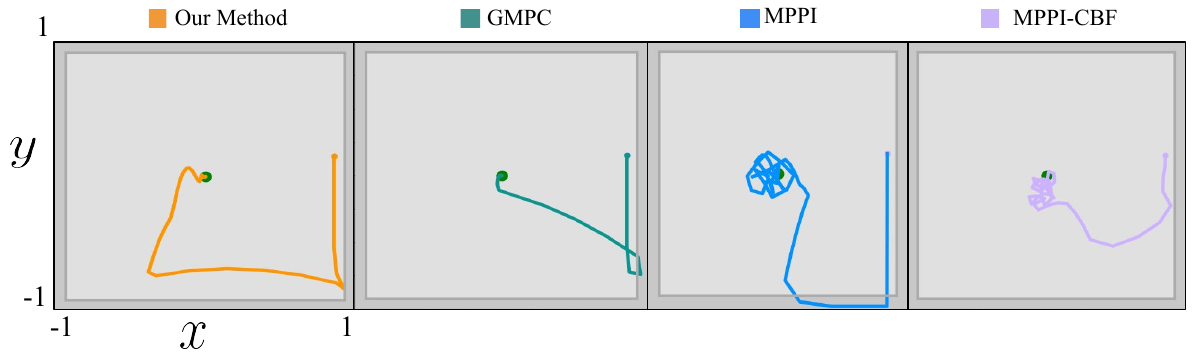}
    \vspace{-2em}
    \caption{Trajectories originating from a common initial state are illustrated, where the dark grey region denotes the walls, floor, and ceiling boundaries. The proposed method successfully \textbf{reaches the goal at $\mathbf{[0,0]^T}$} (green) while \textbf{ensuring collision avoidance}. In contrast, both MPPI and GMPC violate safety constraints, and MPPI-CBF, though safe, fails to converge to the goal, instead oscillating around it and incurring a higher performance cost.}
    \label{fig: MVC_Trajectories}
    \vspace{-1.5em}
\end{figure*}

In this case study, we consider a \textbf{3D Dubins Car} governed by the following dynamics:
\begin{align}
    \dot{x} = v \cos\theta,~~\dot{y} = v \sin\theta,~~ 
    \dot{\theta} = u_1,
\end{align}
where $(x, y, \theta)$ represent the position and heading of the vehicle, $v$ is a fixed forward velocity, and $u_1$ is the control input. The objective is to drive the system from a randomly initialized state to a predefined goal location in the $xy$-plane while avoiding a circular obstacle over a finite time horizon of $2$ seconds. The MPC operates with a time step of $0.05$ seconds and a planning horizon of $20$ steps.

The running cost and terminal cost in Problem~\ref{prob:3} is defined as: $\norm{[x, y]^T - [x_g, y_g]^T}$, where $[x_g, y_g]^T$ denotes the goal position. The safety/state constraint $l$ described in Problem~\ref{prob:1} is formulated as $\norm{[x, y]^T - [x_o, y_o]^T} - r$ where $[x_o, y_o]^T$ is the center of the obstacle and $r$ is its radius.

To enforce safety, we employ a Higher-Order Control Barrier Function (HOCBF)~\cite{9516971, 9029455}, defined as:
\begin{equation}
    h(x,t) = 2xv \cos\theta + 2y v \sin\theta + \alpha(x^2 + y^2 - r^2),
\end{equation}
where $\alpha$ is a positive scalar that controls the rate of convergence to the safe set.

As shown in Figure~\ref{fig:oa plots}, our method successfully navigates the system to the goal while maintaining a safe distance from obstacles, even under challenging initial conditions. In comparison, both MPPI and Gradient-Based MPC (GMPC) fail to enforce safety, resulting in constraint violations. While MPPI-CBF successfully guarantees safety via barrier constraint enforcement, its resulting behavior tends to be overly conservative. This highlights the benefit of employing gradient-based planning to design reference controllers for the CBF-QP filter, enabling improved performance without compromising safety.

A quantitative comparison in Table~\ref{comp table} further highlights the advantage of our approach. Notably, MPPI-CBF incurs a mean cost that is $47.6\%$ higher than our method, indicating reduced performance. Additionally, it exhibits a greater computational burden. Both GMPC and MPPI demonstrate lower safety rates, underscoring the necessity of integrating formal safety guarantees through CBFs. While GMPC offers improved computational efficiency suitable for real-time applications, its lack of safety enforcement limits its practicality. In contrast, our approach provides a principled balance between safety and task performance, addressing the shortcomings of other baselines.

\subsection{Quadrotor Navigation within a Closed Room Setup}

\noindent
In this case study, we consider a \textbf{6D planar quadrotor} system \cite{7525253} with state vector $[x, z, \theta, \dot{x}, \dot{z}, \dot{\theta}]$. The dynamics are governed by:
\begin{equation}
\begin{aligned}
\dot{x} = \dot{x},\quad &\dot{z} = \dot{z},\quad \dot{\theta} = \dot{\theta},\quad m\ddot{x} = F\cos\theta, \\
m\ddot{z} &= F\sin\theta - mg,\quad J\ddot{\theta} = M,
\end{aligned}
\end{equation}
where $x$, $z$, and $\theta$ denote the position in the $x$–$z$ plane and pitch, and $\dot{x}$, $\dot{z}$, $\dot{\theta}$ are the corresponding velocities. $F$ and $M$ represent thrust and torque control inputs, while $m$ and $J$ denote mass and moment of inertia (set to 1 in our simulations). The quadrotor is required to navigate to a goal state from an initial state in the $x$–$z$ plane while avoiding collisions with boundaries (walls, ceiling, and floor) over a 3-second horizon. The MPC time step is 0.05 seconds with a planning horizon of 20 steps.

The running cost, defined in Problem~\ref{prob:3}, is $r(x) = \|[x, z]^T - [x_g, z_g]^T\|$, where $[x_g, z_g]^T$ denotes the target position. The state constraint from Problem~\ref{prob:1} is captured by $\min(0.9-z,~0.9+z,~0.9-x,~0.9+x)$, corresponding to proximity to boundaries. To handle these constraints, we formulate four prioritized CBF-QPs—each addressing a single constraint. The QPs are ordered based on the minimum distance to the respective boundary at each time step. The solution from each QP is recursively passed as the nominal controller to the subsequent QP, thereby forming a safety-prioritized filtering mechanism.
The derived Higher-Order Control Barrier Function (HOCBF) conditions corresponding to each constraint are:
\begin{equation}
\begin{aligned}
C_{1} &= -F\omega\cos\theta - 3\alpha(g - F\sin\theta) - 3\alpha^2\dot{z} + \alpha^3(0.9 - z), \\
C_{2} &= F\omega\cos\theta + 3\alpha(F\sin\theta - g) + 3\alpha^2\dot{z} + \alpha^3(0.9 + z), \\
C_{3} &= F\omega\sin\theta - 3\alpha F\cos\theta - 3\alpha^2\dot{x} + \alpha^3(0.9 - x), \\
C_{4} &= -F\omega\sin\theta + 3\alpha F\cos\theta - 3\alpha^2\dot{x} + \alpha^3(0.9 + x).
\end{aligned}
\end{equation}

As illustrated in Figure~\ref{fig: MVC_Trajectories}, our method successfully drives the system to the goal while maintaining strict adherence to safety constraints, even from a challenging initial position close to the right wall.
In contrast, both MPPI and Gradient-Based MPC undergo safety violations, thereby confirming their inability to ensure safety. Although MPPI-CBF successfully enforces safety constraints, it demonstrates degraded performance due to its inability to converge to the desired goal state, thereby incurring a higher cumulative cost. These observations validate the advantage of leveraging a gradient-based reference controller within the CBF-QP framework to enable the synthesis of safe and performant controllers.

Table~\ref{comp table} shows that both GMPC and MPPI achieve lower safety rates compared to our framework, highlighting the importance of incorporating formal safety guarantees via CBFs. MPPI-CBF, while ensuring safety, incurs a mean cost approximately $110.6\%$ higher than our method and entails greater computational overhead, reflecting reduced performance efficiency. These trends align with observations from the unicycle case study, supporting the claims of generalizability of our approach to higher-dimensional systems. Notably, the performance gap between sample-based and gradient-based techniques widens in the quadrotor setting, where our method exhibits less degradation with increasing dimensionality. Furthermore, the computation time for gradient-based methods scales approximately linearly—doubling from 3D to 6D—whereas sample-based methods experience a $3$–$4$ times increase, indicating superlinear growth. These findings confirm that our framework offers a scalable solution for synthesizing real-time controllers that are both safe and performant, even in complex, high-dimensional environments.

\section{Conclusion}
\label{Conclusion}
\noindent
This work presents a novel two-stage control framework that integrates gradient-based Model Predictive Control (MPC) with Control Barrier Function (CBF)-based safety filtering to jointly optimize performance and safety. By relaxing safety constraints in the optimization stage and subsequently enforcing them via a CBF-QP, the method overcomes the limitations of both overly conservative safety filters and intractable state-constrained optimal control formulations. Empirical validation on two case studies confirms the framework's ability to synthesize controllers that are not only provably safe but also high-performing, demonstrating strong generalizability and computational efficiency across complex control tasks. Future research will investigate learning-based approaches for constructing formally verified CBFs, aiming to enhance scalability to more complex systems~\cite{tayal2024learning, tayal2025cpncbfconformalpredictionbasedapproach}. We will also apply our method to other high-dimensional autonomous systems and systems with unknown dynamics.

\bibliographystyle{IEEEtran}
\bibliography{refs.bib}

\end{document}